\documentclass[useAMS,usenatbib]{mn2e}

\title[Radio constraints on the volume filling factors of AGN winds]{Radio constraints on the volume filling factors of AGN winds}
\author[A. J. Blustin \& A. C. Fabian]{A. J. Blustin$^{1}$\thanks{E-mail: ajb@ast.cam.ac.uk (AJB)} and A. C. Fabian$^{1}$ \\
$^{1}$Institute of Astronomy, University of Cambridge, Madingley Road, Cambridge CB3 0HA \\
}
\begin{document}

\date{Accepted 2009 April 1.  Received 2009 April 1; in original form 2009 Februry 19}

\pagerange{\pageref{firstpage}--\pageref{lastpage}} \pubyear{2009}

\maketitle

\label{firstpage}

\begin{abstract}
The calculation of mass outflow rates of AGN winds is of great importance in understanding the role that such winds play in AGN-galaxy feedback processes. The mass outflow rates are, however, difficult to estimate since the volume filling factors of the winds are unknown. In this paper, we use constraints imposed by the observed radio emission to obtain upper limits to the volume filling factors of wind components in certain nearby AGN. We do this by predicting the 1.4~GHz radio flux densities emitted by those components, assuming a uniform wind, and then comparing these with the observed flux densities for each AGN at this frequency. We find that the upper limits to the volume filling factors are in the range $10^{-4}-0.5$.
\end{abstract}


\section{Introduction}
\label{introduction}

Active galactic nuclei (AGN) drive photoionized winds which produce absorption features in X-ray and UV spectra. The X-ray absorbing part of the wind has historically been termed the X-ray `warm absorber' in the literature; `warm' since the absorption is from gas that is partially ionized, rather than fully stripped or neutral. The winds can contain plasma at a range of ionization levels and ouflow speeds, depending upon the type of AGN; the winds in nearby Seyfert galaxies may be outflowing at a few hundred km~s$^{-1}$, with more highly ionized components outflowing at more than a thousand km~s$^{-1}$ \citep[see~e.g.][]{blustin2005,mckernan2007,reeves2004}. In broad absorption line quasars (BALQSOs), even the low-ionization rest-frame UV absorbing part of the outflow will be moving at thousands of km~s$^{-1}$ along our line of sight \citep[e.g.][]{krolik1999}.

There are as yet no final conclusions on exactly where these winds are driven from. The emerging picture seems to be that different components of AGN winds originate from (or simply become observable at) different locations within the AGN environment. In Seyferts, the fastest and most highly ionized wind components, which are only observable through their absorption features above $\sim$7~keV, may be part of an accretion disc wind \citep[][~and~references~therein]{schurch2008}. Much of the more lowly-ionized and slower-moving plasma, which produces absorption lines in the soft X-ray and/or UV bands, seems to be far away from the central engine, perhaps associated with the Narrow Line Region (NLR) outflow \citep[e.g.][]{kinkhabwala2002}, and may originate in material driven off the putative dusty torus \citep[e.g.][]{blustin2005,mckernan2007,dorodnitsyn2008,chelouche2008}.

AGN winds are important as they are one of the most probable feedback mechanisms that could mediate the co-evolution of supermassive black holes and their host galaxy bulges \citep[e.g.][]{silk1998}. In order to understand their possible role in this process, we need to know how much mass and therefore energy they transport, and when in cosmological history this occurs. Unfortunately, the mass outflow rates of AGN winds are difficult to estimate, since the volume filling factors of the winds are unknown; a new observational window is required. 

Radio emission by thermal bremsstrahlung has long been used as a diagnostic of stellar winds \citep{panagia1975,wright1975}, but has only recently begun to attract attention in the context of winds in radio-quiet AGN \citep{gallimore2004,blundell2007}. In this paper, we estimate the radio flux density due to the ionized winds in five nearby AGN, and, by comparison with the observed values, establish upper limits to the volume filling factors of the wind components with the highest mass outflow rates.

We assume a cosmology with H$_{0}$=70~km~s$^{-1}$~Mpc$^{-1}$, $\Omega_{\rm M}$=0.3, and $\Omega_{\Lambda}$=0.7.

\section{Method}

\subsection{The AGN sample}

In selecting AGN for this study, we used the following criteria: the sources needed to have sufficiently detailed published analyses of their warm absorber properties, based on recent high-quality data, for us to extract the parameters needed for our analysis (ionization parameter, outflow speed), and they also had to have a point-like radio morphology in their 1.4~GHz NRAO VLA Sky Survey \citep[NVSS;][]{condon1998} images obtained via NED\footnote{NASA Extragalactic Database, http://nedwww.ipac.caltech.edu/}, so the nuclear component has minimal contribution from radio emission originating in star formation in the host. 

We checked the Piccinotti sample of X-ray bright AGN \citep{piccinotti1982} against these requirements, also searching through the literature for other well-studied AGN which were not in the original Piccinotti sample. Four objects matched our criteria: NGC~3516, IRAS~13349+2438, MR~2251-178 and MCG-6-30-15. We also included NGC~3783, which has probably the best-studied warm absorber of all AGN, but has a slightly asymmetric and extended 1.4~GHz morphology. The relevant properties of the warm absorbers of the five sources are listed in Table~\ref{wa_properties_results}. 

NGC~3516 is a Seyfert 1 galaxy with a high column, variable warm absorber \citep[e.g.][]{guainazzi2001,netzer2002,kraemer2002,turner2005}. The most recent work on this source claims that the absorber contains components from both a disc wind and from much more distant gas, proposing a model involving four separate absorber components \citep{turner2008}. A variable partial covering component is used to reproduce spectral curvature that has been modelled by other authors \citep[e.g.][]{markowitz2008} as a relativistically-broadened iron line from an accretion disc. In this analysis, we use the absorber parameters provided by \citet{turner2008} as representative of the `maximal' extent of the AGN wind contribution. 

IRAS~13349+2438 is a type-1 radio-quiet quasar whose warm absorber was first studied with high-resolution X-ray grating spectroscopy by \citet{sako2001}. Those authors found that the absorber contained both a low and a high ionisation component; the ionisation structure has been studied in greatest detail by \citet{holczer2007}, who fit absorption columns of individual ions detected in the spectrum and use this to derive an Absorption Measure Distribution (AMD; $d$log(N$_{\rm H}$)/$d$log($\xi$)) for the absorber. Here, $\xi$ is the ionization parameter, defined as 

   \begin{equation}
   \xi = \frac{L_{\rm ion}}{n r^{2}} \,,
\label{xi_definition}
   \end{equation}

where $L_{\rm ion}$ is the 1-1000~Ryd (13.6~eV-13.6~keV) ionizing luminosity, $n$ is the gas density and $r$ is the distance from the ionizing source. \citet{holczer2007} also find that the absorber has two principal ionisation regimes, which they interpret as evidence of thermal instability at the `missing' ionisation levels. For the purposes of our analysis, we use the four bins of their AMD to represent four separate absorber components.

MR~2251-178 is another type-1 radio-quiet quasar, and was the first AGN found to have a soft X-ray ionized absorber \citep{halpern1984}. \citet{kaspi2004} published the first X-ray grating spectrum of this source, finding that the absorber had two or three components. More recently, \citet{gibson2005} have found evidence of another absorber phase with high ionization and very high velocity (v$\sim$12700~km~s$^{-1}$). Here we use the parameters of the \citet{kaspi2004} two-phase model, as quoted by \citet{blustin2005} who estimated $\xi$-values for the ionisation parameter from the original $U$-values given by \citet{kaspi2004}, plus the high-velocity phase discovered by \citet{gibson2005}.

MCG-6-30-15, a narrow-line Seyfert 1, is not detected in the NVSS, but has previously been found to be an unresolved source in VLA imaging \citep{ulvestad1984}. It is the prototypical relativisically-broadened iron K$\alpha$ line source \citep{tanaka1995}, which also has a well-studied warm absorber \citep[e.g.][]{sako2003,turner2004}. This source has been and remains the focus of much debate about the precise contribution of relativistic reflection versus partial covering and/or ionized absorption \citep[e.g.][]{lee2001,miller2008}. Most recently, \citet{miller2008} have presented a model in which all of the X-ray spectral curvature and variability is attributed to absorption effects. Their warm absorber model consists of five components, of which the first two have velocity shifts consistent with zero, the third is outflowing at high speed, and components 4 and 5 would not be expected to produce any detectable narrow features with which to constrain any outflow velocity. We use the \citet{miller2008} warm absorber parameters in our analysis, since, as with NGC~3516, they represent a maximal set of properties for the X-ray absorbing wind. 

The Seyfert 1 NGC~3783 possesses the most extensively studied warm absorber \citep[see~e.g.][]{blustin2002,krongold2003,behar2003,netzer2003,goncalves2006}. We use here the model published by \citet{netzer2003}, in which there are three ionization states, each containing two velocity components. We use $\xi$-values converted from the original $U$ by \citet{blustin2005}. 

\begin{table*}
 \centering
 \begin{minipage}{176mm}
\caption{Properties of the X-ray absorbing photoionized winds (warm absorbers) of the five AGN in our sample, as given in the quoted references (values obtained from elsewhere are referenced in the footnotes), with predicted and observed radio flux densities, and upper limits to the volume filling factors where obtainable: source name, redshift from NED (NASA Extragalactic Database: http://nedwww.ipac.caltech.edu/), and reference for warm absorber model; warm absorber component number identifying each ionization phase; log $\xi$, log ionisation parameter in erg~cm~s$^{-1}$; log $N_{\rm H}$, log total equivalent Hydrogen column density of absorber in cm$^{-2}$; $f_{\rm cov}$, absorber covering fraction; $v$, measured outflow velocity shift of absorber phase in km~s$^{-1}$; $v_{\rm turb}$, turbulent velocity width ($\sigma$) of absorber phase in km~s$^{-1}$; log $T$, log electron temperature of absorber in K, obtained from {\sc spex} xabs model unless otherwise stated (see section~\ref{calculation}); $\dot M_{4\pi}$, mass outflow rate of the wind component assuming a uniform spherically-symmetric wind in M$_{\rm \odot}$~yr$^{-1}$; $S_{\nu,4\pi}$, predicted 1.4~GHz (observed frame; at 1.5~GHz in the case of MGC-6-30-15) radio flux densities, assuming a uniform spherically-symmetric wind, in units of mJy (see section~\ref{calculation}); $S_{\nu,obs}$, observed frame 1.4~GHz flux density in mJy from the NVSS source catalogue \citep{condon1998} and, for MCG-6-30-15, at 1.5~GHz from \citet{ulvestad1984}; $C_{\rm v}$, the upper limits to the volume filling factor if $\Omega$ is in the range 2$\pi-$1.6 (see section~\ref{results}).}
\label{wa_properties_results}
  \begin{tabular}{@{}llllllllllll@{}}
  \hline
Source & Phase & log $\xi$ & log $N_{\rm H}$ & $f_{\rm cov}$ & $v$ & $v_{\rm turb}$ & log $T$ & $\dot M_{4\pi}$ & $S_{\nu,4\pi}$ & $S_{\nu,obs}$ & $C_{\rm v}$  \\
  \hline
NGC~3516                              & 1 & -2.43 & 21.4 & 1    & 200\footnote{Taken as the velocity of the strongest UV absorption lines as in \citet{turner2005}.} & 200  & 4.0 & $1.1\times10^{5}$ & $3.1\times10^{5}$  & 31.3 & 0.0002$-$0.0006 \\
z=0.008836\footnote{\citet{keel1996}} & 2 & 0.25  & 20.7 & 1    & 200\footnote{Taken as the velocity of the strongest UV absorption lines as in \citet{turner2005}.}   & 200  & 4.3 & 220              & 87                &      & 0.07$-$0.3 \\
\citet{turner2008}                    & 3 & 2.19  & 23.3 & 0.45 & 1575  & 1210 & 5.7 & 20               & 0.28              &      &  \\
                                      & 4 & 4.31  & 23.4 & 1    & 1000  & 2124 & 8.1 & 0.096            & $5.3\times10^{-4}$ &      &   \\
 & & & & & & & & & & & \\
IRAS 13349+2438                      & 1 & -0.75 & 21.4 & 1    & 300   & 640  & 4.1\footnote{Temperatures from \citet{holczer2007}.} & $6.9\times10^{4}$ & 590                & 19.6 & 0.01$-$0.05  \\
z=0.107641\footnote{\citet{kim1995}} & 2 & 0.25  & 21.5 & 1    & 300   & 640  & 4.3 & $6.9\times10^{3}$ & 28                 &      & 0.1$-$0.5  \\
\citet{holczer2007}                  & 3 & 2.25  & 21.5 & 1    & 300   & 640  & 5.3 & 69               & $7.1\times10^{-2}$ &      &   \\
                                     & 4 & 3.25  & 21.5 & 1    & 300   & 640  & 5.8 & 6.9              & $3.6\times10^{-3}$ &      &   \\
 & & & & & & & & & & &  \\
MR 2251-178                               & 1 & 0.68  & 20.3 & 0.8  & 300   & 140  & 4.4 & $8.1\times10^{3}$ & 110                & 16.2 & 0.04$-$0.1 \\
z=0.063980\footnote{\citet{bergeron1983}} & 2 & 2.9   & 21.5 & 0.8  & 300   & 140  & 6.6 & 49               & 0.19               &      &   \\
\citet{kaspi2004}                         & 3\footnote{The properties of this phase are taken from \citet{gibson2005}.} & 3.05  & 22.8 & 1    & 12700 & 3400 & 7.0 & $1.5\times10^{3}$ & 0.12               &      &   \\
 & & & & & & & & & & &  \\
MCG-6-30-15                             & 1 & 2.64  & 21.4 & 1    & 0     & 100  & 6.2 & 0                & 0.10 & 1.7  &   \\
z=0.007749\footnote{\citet{fisher1995}} & 2 & 0.25  & 20.2 & 1    & 0     & 100  & 4.3 & 0                & 130  &   & 0.006$-$0.02  \\
\citet{miller2008}                      & 3 & 3.85  & 22.3 & 1    & 1800  & 500  & 8.0 & 0.52             & $2.9\times10^{-3}$ &      &  \\
                                        & 4 & 1.83  & 23.5 & 1    & 0     & 100  & 5.2 & 0                & 1.1 &      &       \\
                                        & 5 & 1.75  & 22.7 & 0.75\footnote{Varies between $\sim$0.5 and 1.} & 0     & 100  & 5.1 & 0                & 1.5 &      &       \\
 & & & & & & & & & & &  \\
NGC 3783                                 & 1 & 1.1 & 21.7 & 1 & 500  & 250 & 4.5 & 210 & 20                 & 43.6 &   \\
z=0.00973\footnote{\citet{theureau1998}} & 2 & 1.1 & 21.5 & 1 & 1000 & 250 & 4.5 & 410 & 20                 &      &   \\
\citet{netzer2003}                       & 3 & 2.3 & 21.8 & 1 & 500  & 250 & 6.2 & 13 & 0.64               &      &   \\
                                         & 4 & 2.3 & 21.6 & 1 & 1000 & 250 & 6.2 & 26 & 0.64               &      &   \\
                                         & 5 & 2.9 & 22.1 & 1 & 500  & 250 & 7.7 & 3.3 & 0.12               &      &   \\
                                         & 6 & 2.9 & 21.9 & 1 & 1000 & 250 & 7.7 & 6.5 & 0.12               &      &   \\
  \hline
\end{tabular}
\end{minipage}
\end{table*}

The temperatures of the individual warm absorber phases are required in order to estimate the radio bremsstrahlung emission (see section~\ref{calculation}). In one case, IRAS~13349+2438, the temperatures are given in the paper from which we took the other parameters, but for the other sources we calculated them using the xabs photoionized absorber model in {\sc spex} 2.00.11 \citep{kaastra1996}. The xabs model is based on {\sc xstar} \citep{kallman1982} grids generated for a given input spectral energy distribution (SED), and takes Thompson scattering into account as well as photoelectric absorption. For each xabs component, the user can specify the absorbing column, ionization parameter, velocity shift, turbulent velocity, covering fraction and elemental abundances (all abundances are assumed to be Solar $-$ \citet{anders1989} $-$ in this analysis). We constructed a SED for each source which was based on the canonical AGN SED of \citet{marconi2004}, following the general procedure of \citet{blustin2008}. The X-ray power-law slopes $\Gamma_{\rm X}$ were taken from the same publications as the warm absorber parameters, except in the case of NGC~3516 where \citet{turner2008} do not quote a continuum slope, so we assume $\Gamma_{\rm X}=1.9$. The X-ray to optical flux ratios were set using $\alpha_{\rm ox}$ values obtained from the literature. The SEDs were normalised to the $L_{\rm ion}$ value quoted by the authors of the respective warm absorber model, if available, or otherwise from \citet{blustin2005}. The values for $L_{\rm ion}$, $\alpha_{\rm ox}$ and $\Gamma_{\rm X}$ are listed in Table~\ref{sed_properties}, and the derived temperatures are in Table~\ref{wa_properties_results}.

\begin{table}
 \centering
 \begin{minipage}{85mm}
  \caption{Key parameters of the assumed SEDs for the four AGN in our sample: source name; log $L_{\rm ion}$, log 1-1000~Ryd ionising luminosity in erg~s$^{-1}$ from \citet{blustin2005} unless otherwise stated; $\alpha_{\rm ox}$, optical-to-X-ray power-law slope \citep{tananbaum1979} from \citet{blustin2005} unless otherwise stated; $\Gamma_{\rm X}$, X-ray power-law slope.}
\label{sed_properties}
  \begin{tabular}{@{}llll@{}}
  \hline
Source & log $L_{\rm ion}$ & $\alpha_{\rm ox}$ & $\Gamma_{\rm X}$  \\
  \hline
NGC~3516        & 43.7\footnote{\citet{turner2008}} & -1.4\footnote{\citet{vasudevan2009}}  & 1.9\footnote{Assumed value, since $\Gamma_{\rm X}$ not quoted by \citet{turner2008}} \\
IRAS~13349+2438 & 45.0 & -1.31 & 1.9\footnote{\citet{holczer2007}} \\
MR~2251-178     & 45.5 & -1.24 & 1.5\footnote{\citet{kaspi2004}} \\
MCG-6-30-15     & 43.7 & -1.41\footnote{\citet{vasudevan2009}} & 2.2\footnote{\citet{miller2008}} \\
NGC~3783        & 44.1 & -1.28 & 1.65\footnote{\citet{netzer2003}} \\
  \hline
\end{tabular}
\end{minipage}
\end{table}

\subsection{Estimating the radio luminosity due to the winds}
\label{calculation}

The original calculations of the radio emission from stellar winds \citep{panagia1975,wright1975} assumed that the wind was optically thick, with the blackbody emission at each point being propagated through bremsstrahlung absorption along the line of sight. The recent paper of \citet{blundell2007}, however, calculated the flux density from an AGN wind by assuming that the emission emerges from the optically thin part of the wind beyond a photospheric radius $r_{\rm ph}$. To see whether the winds in our AGN fall within this regime, we used Equations 3 and 4 of \citet{blundell2007} to estimate $r_{\rm ph}$ for each absorber phase. We also made a rough estimate of the optical depth of each phase using the approximation

\begin{eqnarray}
   \tau &\sim& K(\nu,T) n^{2} r \nonumber\\
        &\sim& K(\nu,T) \frac{L_{\rm ion}^{2}}{\xi^{2} r^{3}} 
\label{tau_estimate}
\end{eqnarray}

where $K(\nu,T)$ is the linear free-free absorption coefficient \citep[see~e.g.][]{rybicki1979}, $n$ is the gas density, and $r$ is the depth of the absorber. We substitute in the expression for the ionization parameter to remove the non-observable $n$, and approximate $r$ as the distance of the absorber from the ionizing source. We estimate this by assuming that the observed velocity of the phase represents its escape velocity. We find that $r_{\rm ph}$ is large for many of the phases (up to kpc), and that the optical depth of most of the phases is $\gg$1. The expressions of \citet{panagia1975} and \citet{wright1975} (hereafter PF and WB; Equations 24 and 8 in those works respectively) are therefore more appropriate for calculating the radio emission of these winds. These two independently derived expressions are similar but not identical; the principal difference is that WB include the free-free Gaunt factor $g_{\rm ff}$ explicitly, whereas PF account for its temperature and frequency dependence as separate terms. Since $g_{\rm ff}$ can be conveniently estimated from Figure~5 in \citet{karzas1961}, we use the WB formulation to estimate the radio emission of the winds in the five AGN in our sample. 

This expression, being derived for use with stellar winds, need to be altered slightly in the case of AGN. It is based on the assumption that the winds are spherically symmetric and uniformly volume-filling, whilst, in reality, AGN winds have some covering factor $\Omega$ and volume filling factor $C_{\rm v}$. The higher redshift of AGN also means that a 1/(1+z) factor is needed to transform the flux to the observed frame. 

The mass outflow rate for a volume-filling, spherically-symmetrical wind, at a distance $r$ from the ionizing source, is

   \begin{equation}
   \dot M_{4\pi} = 4 \pi r^2 n \mu m_{\rm p} v \,,
   \end{equation}

where $n$ is the gas density, $\mu$ is the mean atomic weight of the gas, $m_{\rm p}$ is the proton mass and $v$ is the outflow speed. Substituting in the expression for the ionization parameter $\xi$ (Equation~\ref{xi_definition}), 

   \begin{equation}
   \dot M_{4\pi} = \frac{4 \pi L_{\rm ion} \mu m_{\rm p} v}{\xi} \,.
\label{mdot_4pi}
   \end{equation}

For an AGN wind, however, we must introduce a volume filling factor $C_{\rm v}$ and a global covering factor $\Omega$, so that 

   \begin{eqnarray}
   \dot M &=& \frac{4 \pi L_{\rm ion} \mu m_{\rm p} v C_{\rm v} \Omega}{\xi} \nonumber\\
          &=& \dot M_{4\pi} C_v \Omega                 \,.
\label{mdot_defn}
   \end{eqnarray}

Including the flux conversion, the expression for radio flux density from a uniform, spherically-symmetric wind, $S_{\nu}$, given by WB is:

   \begin{equation}
   S_{\nu,4\pi} = 23.2 \left(\frac{\dot M_{4\pi}}{\mu v}\right)^{\frac{4}{3}} \nu^{\frac{2}{3}} \gamma^{\frac{2}{3}} g_{\rm ff}^{\frac{2}{3}} \bar Z^{\frac{4}{3}} d_{\rm kpc}^{-2} (1+z)^{-1} \, Jy,
\label{s_nu_WB}
   \end{equation}

where $\dot M_{4\pi}$ is M$_{\rm \odot}$~yr$^{-1}$, $\nu$ is the rest-frame frequency corresponding to the observed radio frequency, $\gamma=n_{\rm e}/n_{\rm i}$ is the ratio of electron density to ion density, $g_{\rm ff}$ is the free-free Gaunt factor, $\bar Z$ is the average ionic charge, and $d_{\rm kpc}$ is the distance to the source in kpc.

Because 

   \begin{equation}
S_{\nu,4\pi} \propto \dot M_{4\pi}^{\frac{4}{3}} \,,
   \end{equation}

then, substituting in Equation~\ref{mdot_defn}, the actual observed flux density $S_{\nu}$ is

   \begin{equation}
S_{\nu} \propto ( C_v \Omega )^{\frac{4}{3}} \dot M_{4\pi}^{\frac{4}{3}} \,.
   \end{equation}

It follows that the product $C_{\rm v} \Omega$ can be expressed as

   \begin{equation}
C_{\rm v} \Omega = \left(\frac{S_{\nu}}{S_{\nu,4\pi}}\right)^{\frac{3}{4}} \,,
\label{cv_omega}
   \end{equation}

so $C_{\rm v} \Omega$ can therefore be estimated using the ratio of the observed radio flux density to that predicted for a spherically-symmetric uniform wind.

We note that most of the absorber phases described by \citet{miller2008} for MCG-6-30-15 are not outflowing. We can predict the radio flux from these, if we assume that the density $n \propto 1/r^2$ and that $\xi$ is constant; we can then make the trivial substitution

   \begin{eqnarray}
\frac{\dot M_{4\pi}}{\mu v} &=& 4 \pi m_{\rm p} n r^2 \nonumber\\
                           &=& 4 \pi m_{\rm p} \frac{L_{\rm ion}}{\xi}
\label{s_nu_v0}
   \end{eqnarray}

into Equation~\ref{s_nu_WB} for the phases with zero outflow velocity.

\section{Results}
\label{results}

We use Equation~\ref{s_nu_WB} to estimate the 1.4~GHz flux density emitted by each wind phase, assuming a uniform, spherically-symmetric wind; the values are listed in Table~\ref{wa_properties_results}, alongside the associated mass outflow rates calculated using Equation~\ref{mdot_4pi}. For the purposes of this calculation, we assume that $\mu=1.23$, $\gamma=1$, $\bar Z=1.4$ and extrapolate $g_{\rm ff}$ from Figure~5 in \citet{karzas1961}. It turns out that, generally, only the phases with the highest mass outflow rates are predicted to produce radio flux densities greater than or equal to the observed values; for these phases, we use Equation~\ref{cv_omega} to obtain an upper limit to $C_{\rm v} \Omega$. 

An exception is MCG-6-30-15, where the highest radio flux is predicted for the non-outflowing phase with the lowest ionisation parameter. In this source, the partially-covering disc wind component (phase 5) could be a feasible source of a large part of the observed radio flux density, \emph{if} it has a volume filling factor of near unity. If, on the other hand, its volume filling factor is comparable to that of phase 2, which must be less than 0.02 (see Table~\ref{wa_properties_results}), then this is unlikely. In the case of NGC~3783, none of the wind phases can produce a significant fraction of the observed radio flux density, although the combined flux of all phases comes close. 

The global covering factor $\Omega$ of AGN winds is uncertain. It is known that $\sim$50\% of nearby Seyfert 1s have a wind \citep{reynolds1997}, implying a covering factor of $2\pi$, which would be entirely possible for an accretion disc wind. It seems increasingly likely, however, that most of the observed absorbing column in the winds is associated with gas at the distance of the torus or NLR, implying that the covering factor cannot be greater than the solid angle of the openings in the torus. The minimum covering factor will therefore be the product of the percentage of AGN that are unobscured (i.e. the fraction of AGN that are type 1; 25\%) and the percentage of type 1 AGN with warm absorbers, giving $\Omega=1.6$. For the purposes of our calculations, we use these two extremes of $\Omega$ to give a range of $C_{\rm v}$.

The overall result is that, for the reasonable range of $\Omega$, the upper limits to $C_{\rm v}$ are of the order of $10^{-4}-0.5$; the figures are listed in Table~\ref{wa_properties_results}.

\section{Discussion and conclusions}
\label{discussion}

We have shown that, at least for some phases of X-ray absorbing winds, the volume filling factors need to be small: of the order of $10^{-4}-0.5$. They may well be lower than this, since we have not taken into account the radio emission from the UV-absorbing part of the outflow, and we have assumed that there is no contribution from the host galaxies, the base of any radio jet, or indeed emission from the accretion disc corona \citep[as~proposed~by][]{laor2008}. Certainly, in the case of NGC~3783 where we do not predict the X-ray absorbing wind to be a feasible source of the observed radio emission, the fact that the radio source is slightly extended may imply that emission from the host galaxy is a significant component. Any future observations of these sources which resolve jet or host galaxy components would allow us to further constrain the volume filling factors.

If the filling factors of the wind phases are small, it is interesting to estimate the radio emission from the plasma filling the rest of the space. A possible scenario is that the relatively low-ionisation phases with measurable filling factors are embedded in a hot medium, outflowing at the same speed, with which they are in pressure equilibrium. In this case, assuming, for example, that $T=10^8$~K for the hot plasma, we find that the radio flux densities due to the confining media are at least two orders of magnitude lower than the observed radio flux densities for each source. 

The radio emission from radio-quiet AGN can be variable on timescales of a month or less \citep{barvainis2005}, which is probably much shorter than the timescale of significant changes to the mass outflow rate in parsec-scale AGN winds. On the other hand, if some fraction of the radio emission originated from a partially-covering disc wind, and if the mass outflow rate and therefore radio flux varied according to how much of the source the wind was covering, then correlated variations in radio flux and absorber partial covering could provide observational support for the existence of such wind components. 

\section*{Acknowledgments}

AJB and ACF acknowledge the support of, respectively, an STFC Postdoctoral Fellowship and the Royal Society. We thank L. Miller for providing extra information about the analysis of \citet{miller2008}. This research has made use of the NASA/IPAC Extragalactic Database (NED) which is operated by the Jet Propulsion Laboratory, California Institute of Technology, under contract with the National Aeronautics and Space Administration.

\label{lastpage}

\end{document}